\documentclass[%
 preprint,
 superscriptaddress,
 amsmath,amssymb,
 aps,
 pra,
]{revtex4-1}

\usepackage{graphicx}
\usepackage{dcolumn}
\usepackage{bm}

\usepackage[T1]{fontenc}
\usepackage[utf8]{inputenc}
\usepackage{braket}
\usepackage{booktabs}
\usepackage{units}

\newcommand{\vect}[1]{\boldsymbol{\mathbf{#1}}}
\newcommand{\hatt}[1]{\mathbf{\hat{#1}}}

\graphicspath{ {../data/} }

\begin{document}

\title{Outcoupling from a Bose-Einstein condensate in the strong-field limit}

\author{Caroline Arnold}
\email{caroline.arnold@cfel.de}
\altaffiliation{Present address: Center for Free-Electron Laser Science, DESY, Notkestrasse 85, 22607 Hamburg, Germany}
\affiliation{Institut für Theoretische Physik, Universität Tübingen, Auf der Morgenstelle 14, D-72076 Tübingen, Germany}

\author{Carola Beck}
\affiliation{Physikalisches Institut der Universität Tübingen, Auf der Morgenstelle 14, D-72076 Tübingen, Germany}

\author{Peter Federsel}
\affiliation{Physikalisches Institut der Universität Tübingen, Auf der Morgenstelle 14, D-72076 Tübingen, Germany}

\author{Malte Reinschmidt}
\affiliation{Physikalisches Institut der Universität Tübingen, Auf der Morgenstelle 14, D-72076 Tübingen, Germany}

\author{József Fortágh}
\affiliation{Physikalisches Institut der Universität Tübingen, Auf der Morgenstelle 14, D-72076 Tübingen, Germany}

\author{Andreas Günther}
\email{a.guenther@uni-tuebingen.de}
\affiliation{Physikalisches Institut der Universität Tübingen, Auf der Morgenstelle 14, D-72076 Tübingen, Germany}

\author{Daniel Braun}
\email{daniel.braun@uni-tuebingen.de}
\affiliation{Institut für Theoretische Physik, Universität Tübingen, Auf der Morgenstelle 14, D-72076 Tübingen, Germany}

\date{\today}

\begin{abstract}
Atoms can be extracted from a trapped Bose-Einstein condensate (BEC) by driving spin-flips to untrapped states. The coherence properties of the BEC are transfered to the released atoms, creating a coherent beam of matter refered to as an atom laser. In this work, the extraction of atoms from a BEC is investigated numerically by solving a coupled set of Gross-Pitaevskii equations in up to three dimensions. The result is compared to experimental data and a semiclassical rate model. In the weak-coupling regime, quantitative agreement is reached between theory and experiment and a semiclassical rate model. In the strong-coupling regime, the atom laser enters a trapped state that manifests itself in a saturation of the rate of out-coupled atoms observed in new experimental data. The semiclassical rate model fails, but the numerical descriptions yield qualitative agreement with experimental data at the onset of saturation.  
\end{abstract}

\maketitle

\section{Introduction}
\label{sect:introduction}
Since Bose-Einstein condensates became experimentally accessible in magnetic and optical traps \cite{anderson95}, output coupling mechanisms have been used to extract atoms from the trap in a controlled way \cite{mewes97}. Typically, radio-frequency (rf) or microwave (mw) magnetic fields are used to drive spin-flips to untrapped states, with the out-coupled atoms carrying the fixed phase relation of the condensate. In close analogy to a photon laser, the resulting coherent matter wave is referred to as an atom laser \cite{robins13}. 

The system of trapped and falling atoms is typically described by a set of coupled Gross-Pitaevskii equations (CGPE) \cite{ballagh97}. In the weak-coupling limit, analytic models have been introduced to describe the out-coupled atom beam \cite{steck98, federsel15, harkonen10, kramer06, kalman16}.

Techniques for the numerical solution of the CGPE have been given in \cite{antoine13, bao04, williams99, bao02}. In \cite{schneider99, harkonen10, steck98}, 1D simulations have been used to compare the model to experimental data. The high-intensity or strong-coupling limit of the atom laser has been studied experimentally in \cite{robins06}. The atom beam can be analyzed on a single-particle level by an appropriate detection scheme \cite{ottl05, federsel15}. If condensate properties can be linked to the characteristics of the atom beam, this will provide a destruction-free way of studying ultracold atomic clouds. 

In this paper, we investigate a microwave-induced atom laser in $^{87}$Rb. It is treated as an effective two-level system and described by a set of coupled Gross-Pitaevskii equations (CGPE). They are solved numerically in up to three dimensions. The numerical results are compared to a rate model and experimental data in the weak-coupling limit given in \cite{federsel15}, as well as new experimental data in the strong-coupling limit. Quantitative agreement is reached in the weak-coupling limit with the three-dimensional simulations. The simulation is extended to the strong-coupling limit, where dynamical processes in the out-coupling region become relevant \cite{robins05}. These are captured by two- and three-dimensional simulations, but not by one-dimensional ones.

\section{Model}
\label{sect:model}

\subsection{Magnetic trap}
We consider a Bose-Einstein condensate of $^{87}\mathrm{Rb}$ atoms in a magnetic trap. Zeeman splitting leads to the separation of the hyperfine magnetic sub-states labeled by $\ket{F, m_F}$, where $\vect F = \vect I + \vect J$ is the vector sum of the nuclear spin $\vect I$ and electron spin $\vect J$. For $^{87}\mathrm{Rb}$ in the $5s_{1/2}$ ground state, $I = 3/2$ and $J = 1/2$. The trapping potential is approximated by a harmonic potential. An offset field $B_\mathrm{off}$ is added to prevent losses by Majorana spin-flips. Gravity is taken into account, where $\hatt e_z$ defines the axis of gravity. The full potential is then given by
\begin{equation}
    \begin{split}
        V(\vect r) = \mathrm{sgn}(g_F) m_F \frac m 2 \left(\omega_x^2 x^2 + \omega_y^2 y^2 + \omega_z^2 z^2\right) 
        + \mu_B m_F g_F B_\mathrm{off} - m g z +F \hbar\omega_0\\
        = V_\mathrm{trap}(\vect r) + \mu_B m_F g_F
        B_\mathrm{off}+F \hbar\omega_0 
        \label{eq:potential}
    \end{split}
\end{equation}
where $g_F = g_J(F(F+1) + J(J+1) - I(I+1))/2F(F+1)$ defines the Landé $g$-factor \cite{metcalf99}. In the $5s_{1/2}$ ground state, $g_J = 2$, thus $g_{F=2} = 1/2, g_{F=1} = -1/2$. Further, $\mu_B$ is the Bohr magneton, and $m$ the $^{ 87 }$Rb mass. The trap frequencies $\omega_{i}$ are given with respect to the $m_F=1$ level. The trapped state is $\ket{F=2, m_F=2}$. The effects of temperature are neglected. We have included the zero-field hyper-fine splitting with $\omega_0 \simeq 2\pi\times \unit[6.835]{GHz}$ in the trap potential.

\subsection{Interaction with radiation}
Electro-magnetic radiation can be used to drive spin-flips between trapped and un-trapped hyperfine sub-levels. We restrict ourselves to the case of a microwave transition $\ket{F=2, m_F=2} \rightarrow \ket{F=1, m_F=1}$, caused by $\vect B_\mathrm{mw}(t) = B_\mathrm{mw} \hatt{e}_\mathrm{mw} \cos(\omega_\mathrm{mw} t)$. The other hyperfine sub-levels are off-resonant and do not have to be taken into consideration. The coupling can thus be treated within the framework of a two-level system. Atoms in the $\ket{F=1, m_F=1}$ hyperfine sub-level are anti-trapped and form an atom beam that can be analyzed by single-atom detection. Experimentally, this has been realized by ion counting following photoionization. The interaction is treated semi-classically, $H_\mathrm{int} = - \vect \mu \vect B_\mathrm{mw}$, where $\vect \mu$ denotes the magnetic dipole moment of the atom. The Rabi frequency of the two-level transition is defined as $\hbar \Omega = g_F \mu_B B_\mathrm{mw} / 2 \bra{F', m_F'} J_\pm \ket{F, m_F}$, where the transition matrix element is included. The microwave is assumed to be correctly polarized, such that a transition with $\Delta F = \Delta m_F = 1$ is possible. 

\subsection{Coupled Gross-Pitaevskii equations}
The system is described by a set of coupled Gross-Pitaevskii equations
(CGPE), following \cite{ballagh97, schneider99}. After transforming to
a rotating frame $\psi_{m_F}(\vect r, t) \rightarrow e^{-i F
    \omega_{\text mw}} t\psi_{m_F}(\vect r, t),$   
and applying the rotating wave approximation, the CGPE for the two-level system read 
\begin{equation}
    i \hbar \frac{\partial}{\partial t}\psi_{1, 2}(\vect r, t)
    = \left(- \frac{\hbar^2}{2m}\nabla^2 + V_{\mathrm{eff},1, 2}(\vect r) \right)\psi_{1, 2}(\vect r, t)
    + \hbar \Omega \psi_{2, 1}(\vect r, t)
    \label{eq:cgpe-rwa-rotframe}
\end{equation}
where the index $i \in \{1, 2\}$ labels the states $\ket{F=2, m_F=2}$ and $\ket{F=1, m_F=1}$, respectively. The effective potential is given by
\begin{equation}
    V_{\mathrm{eff}, i} = V_{\mathrm{trap}, i} + m_{F, i} \hbar {\Delta_i} + g_\mathrm{3D}|\psi(\vect r, t)|^2,
    \label{eq:cgpe-effective-potential}
\end{equation}
where $|\psi(\vect r,t)|^2 = \sum_{i=1}^2 |\psi_i(\vect r,t)|^2$. The
detuning frequency for a hyperfine sub-level with $|m_F|=1$ to the
center of the trap is given by $\hbar {\Delta_i} = \hbar
{(\omega_0-\omega_{\text mw})} + g_F \mu_B
B_\mathrm{off}$, see \cite{schneider99}. The 
inter-atomic coupling constant is given by $g_\mathrm{3D} = 4 \pi
\hbar^2 a N / m$, with the scattering length $a = 110 a_0$, mass $m =
87 m_p$, and the atom number $N$. Here, $a_0$ is the Bohr radius and
$m_p$ the proton mass. The normalization is chosen such that $\int
\mathrm{d}\vect r\, |\psi(\vect r, t)|^2 = 1$. Lower-dimensional
modeling is achieved by requiring that the chemical potential, in
Thomas-Fermi approximation, be the same across dimensions
\cite{williams99}. The corresponding coupling constants in 1D and 2D
are then given by  
\begin{gather}
    g_\mathrm{1D} = 
    \frac{1}{2\pi}\left(\frac{125}{9}\right)^{1/5}\bar{\sigma}_0^{-8 / 5} (a N)^{-2 / 5} \left(\frac{\bar\omega}{\omega_z}\right) |m_{F, \mathrm{trap}}|^{2/5} g_\mathrm{3D},
    \label{eq:g1d}
    \\
    g_\mathrm{2D} =
    \frac{15^{4/5}}{16} \left(\frac{\bar \omega}{\omega_{xy}}\right)^2 (a N)^{-1/5}\bar\sigma_0^{-4/5}|m_{F, \mathrm{trap}}|^{1/5} g_\mathrm{3D},
\end{gather}
respectively, where $\bar \omega = (\omega_x \omega_y \omega_z)^{1/3}$ denotes the geometrically averaged trap frequency, $\bar \sigma_0 = \sqrt{\hbar / m \bar \omega}$ the corresponding oscillator length, and $m_{F, \mathrm{trap}}$ the magnetic quantum number of the trapped state. 

\subsection{Outcoupling}
\label{sect:model:outcoupling}
Energy conservation restricts the transition to the crossing point of the effective potentials given in Eq.~\eqref{eq:cgpe-effective-potential}. These can be shifted by adjusting the detuning frequency relative to the trapped state. As the condensate is displaced from the trap minimum by gravity, centering at the gravitational sag $z_\mathrm{sag} = g / |m_F| \omega_z^2$, a non-zero detuning frequency $\Delta_\mathrm{res}$ is required for output coupling. Maximum outcoupling is achieved when the resonant point matches the gravitational sag. Power broadening has to be taken into account \cite{robins06}, the resonant frequency range is then given by $\Delta_\mathrm{res} \pm \frac 1 2 \Omega$.  
In the weak-coupling limit, the atoms in the un-trapped state leave the resonant area under the influence of gravity. For Rabi oscillations to take place, it is required that $t_\mathrm{res} > t_\mathrm{\Omega} = 2 \pi / \Omega$, where $t_\mathrm{res}$ is the time spent in the resonant range. During the Rabi oscillations, atoms in the anti-trapped state go back to the trapped state before they leave the resonant range. The intensity of the atom laser is thus reduced, and the system enters a bound state as described by \cite{robins05}. 

\subsection{Rate model}
In \cite{federsel15}, microwave outcoupling from both thermal clouds and BEC has been described quasiclassically. In the weak-coupling limit, outcoupling rates are given by 
\begin{equation}
    \Gamma(\omega) = \frac{\pi\Omega^2}{2} \sqrt{\frac{\lambda\hbar}{2 m \omega_z^2}} \frac{n[z(\omega)]}{\sqrt \omega}
    \label{eq:rate-model}
\end{equation}
where $\omega$ denotes the full detuning from the trap center, $\lambda = 1 / (1 - g_F' m_F'  / g_F m_F )$ is a dimensionless parameter depending on the interacting hyperfine sub-levels. The integrated line density at the point of resonance for a given detuning frequency, $n[z(\omega)]$, in the Thomas-Fermi limit is given by 
\begin{equation}
    n (z) = \frac{\mu \pi R_x R_y}{2 g_\mathrm{3D}} \max \left[0, 1 - \frac{\left(z - z_\mathrm{sag}\right)^2}{R_z^2} \right]^2
    \label{eq:integrated-line-density}
\end{equation}
where $\mu$ refers to the chemical potential in Thomas-Fermi approximation and $R_i$ denotes the Thomas-Fermi radius along the respective axis. While the line density derived from the Thomas-Fermi approximation in 1D takes the shape of an inverse parabola, the integrated line density follows a squared inverse parabola.  

\section{Numerical solution}
\label{sect:numerical}

The dimensionless CGPE are given by 
\begin{equation}
    i \frac{\partial}{\partial t} \psi_{1, 2}(\vect r, t) = \left(-\frac 1 2 \nabla^2 + V_\mathrm{eff, 1, 2}(\vect r)\right)\psi_{1, 2}(\vect r, t) + \frac{\Omega}{\omega_z}\psi_{1, 2}(\vect r, t)
    \label{eq:dimless-cgpe}
\end{equation}
where the typical scales are the oscillator length $\sigma_0 = \sqrt{\hbar / m \omega_z}$, the time $t_0 = 1 / \omega_z$ and the energy $\varepsilon_\mathrm{char} = \hbar \omega_z$. The dimensionless effective potential is given by
\begin{equation}
    \begin{split}
        V_{\mathrm{eff}, 1, 2} = \frac 1 2 \mathrm{sgn}(g_F)m_F \left(\frac{\omega_x^2}{\omega_z^2} x^2 + \frac{\omega_y^2}{\omega_z^2} y^2 + z^2\right) \\
        + m_F \frac{\Delta}{\omega_z} - \frac{g}{\sigma_0 \omega_z^2} + \frac{g_\mathrm{3D}}{\hbar \omega_z \sigma_0^3}|\psi(\vect r)|^2
    \end{split}
    \label{eq:dimless-effective-potential}
\end{equation}

The CGPE were solved numerically in up to three dimensions by the symmetrized split-step Fourier method. The formal solution of Eq.~\eqref{eq:dimless-cgpe} is split according to the Strang splitting \cite{bao03}

\begin{equation}
    e^{-i (T + V + W) \tau} = e^{-i T \tau / 2} e^{-i V \tau} e^{-i W \tau} e^{-i T \tau / 2} + \mathcal{O}(\tau^3)
    \label{eq:strang-splitting}
\end{equation}

where each part can be solved analytically. Since, with the given potential operators, $[V_i, V_j] = [T_i, T_j] = 0$, the approach can easily be generalized to higher dimensions. Space is discretized on a mesh with mesh size $h = 1 / 16$, and for the time step $\tau=10^{-4}$ is chosen. An imaginary absorbing potential is added below the detection height to prevent unwanted reflection at the lower end of the grid \cite{antoine13}. The ground state is obtained by propagation in imaginary time \cite{bao04}. 


The ion count rate (ICR) at the detection height $z_\mathrm{det}$, situated below the trap, is calculated via the probability current $j = \frac \hbar m |\psi|^2 \partial_z \varphi(z)$, $\mathrm{ICR} = N \eta j(z_\mathrm{det})$, where $\varphi$ is the phase of the wave function and the detection efficiency is given by $\eta$. 

\section{Results}
\label{sect:results}

\begin{table}
    \centering
    \caption{Experimental parameters as given in \cite{federsel15} and used in the simulation. Atom number and detection efficiency are given for the spectral response data and, in round brackets, for the sensitivity data. Note that, in the latter case, the detection efficiency varies in the weak-coupling (wc) and strong-coupling (sc) data. Trap frequencies are given with respect to the $|m_F| = 1$ level.}
    \label{tab:experiment-parameters}
    \begin{tabular}{lcc}
        \toprule
        Quantity    &   Symbol  &   Value   \\
        \hline 
        Atom number &   $N$ &   10000 (  8200 ) \\
        Detection efficiency    &   $\eta$  &   0.24    ( 0.073 (wc) / 0.007 (sc) ) \\
        Trap frequency  &   $\omega_x$  &   $\unit[1 / \sqrt 2 \times 2 \pi \times 16]{Hz}$ \\
        Trap frequency  &   $\omega_y$  &   $\unit[1 / \sqrt 2 \times 2 \pi \times 85]{Hz}$ \\
        Trap frequency  &   $\omega_z$  &   $\unit[1 / \sqrt 2 \times 2 \pi \times 70.75]{Hz}$ \\
        \bottomrule
    \end{tabular}
\end{table}

\begin{figure}
    \centering
    \includegraphics{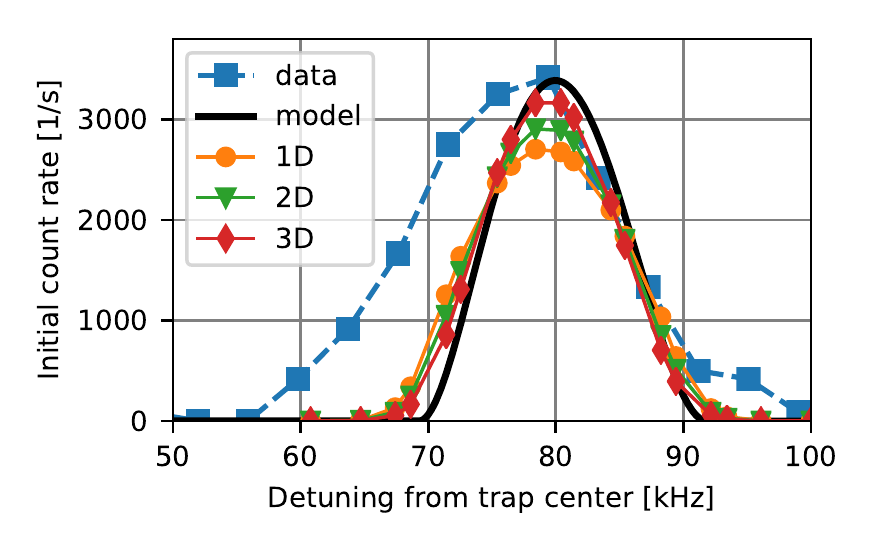} 
    \caption{Spectral response. Experimental data (blue squares) obtained by sweeping the detuning frequency at a rate of $\unit[1.3]{MHz/s}$ and at fixed Rabi frequency in the weak-coupling limit $(\Omega=\unit[2\pi\times 83]{s^{ -1}})$ is compared to the rate model, as well as to the numerical solution of the CGPE in 1D, 2D, and 3D, respectively. The simulations and the rate model were conducted pointwise at fixed detuning frequencies.}
    \label{fig:spectral-response}
\end{figure}

\begin{figure}
    \centering
    \includegraphics{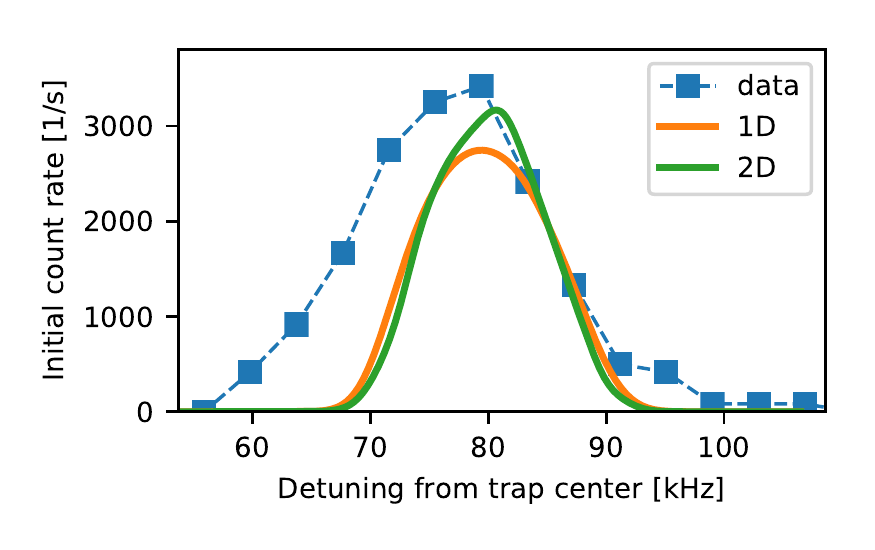}
    \caption{Spectral response. Experimental data (same as in Fig.~\ref{fig:spectral-response}) is compared to the numerical solution of the CGPE in 1D and 2D by sweeping the resonant point through the condensate with a rate of $\unit[1.3]{MHz/s}$. Compared to Fig.~\ref{fig:spectral-response}, the detuning frequency is here adapted with the sweeping rate.}
    \label{fig:spectral-response-sweep}
\end{figure}

\begin{figure}
    \centering
    \includegraphics{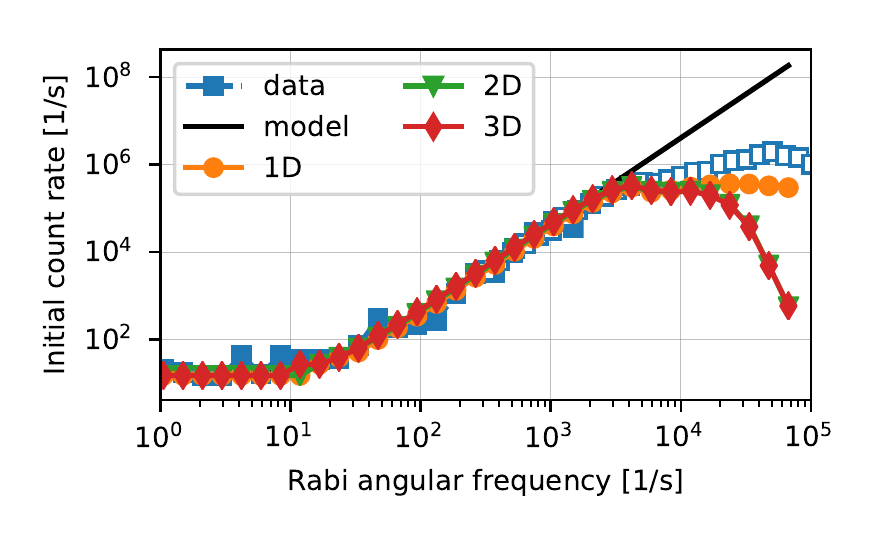}
    \caption{Sensitivity. Experimental data \textit{(blue squares)} is obtained at the detuning frequency for maximum outcoupling with a detection efficiency of $\eta = 0.073$ in the weak-coupling limit \textit{(filled squares)} and $\eta = 0.007$ in the strong-coupling limit \textit{(empty squares)}. The latter data has been rescaled to the detection efficiency of the former. The experimental data is compared to the semiclassical rate model and the numerical solution of the CGPE in 1D, 2D, and 3D \textit{(colored lines)}, respectively. The background count rate of $\unit[15]{s^{-1}}$ is taken into account. }
    \label{fig:sensitivity}
\end{figure}

\begin{figure*}
    \centering
    \includegraphics{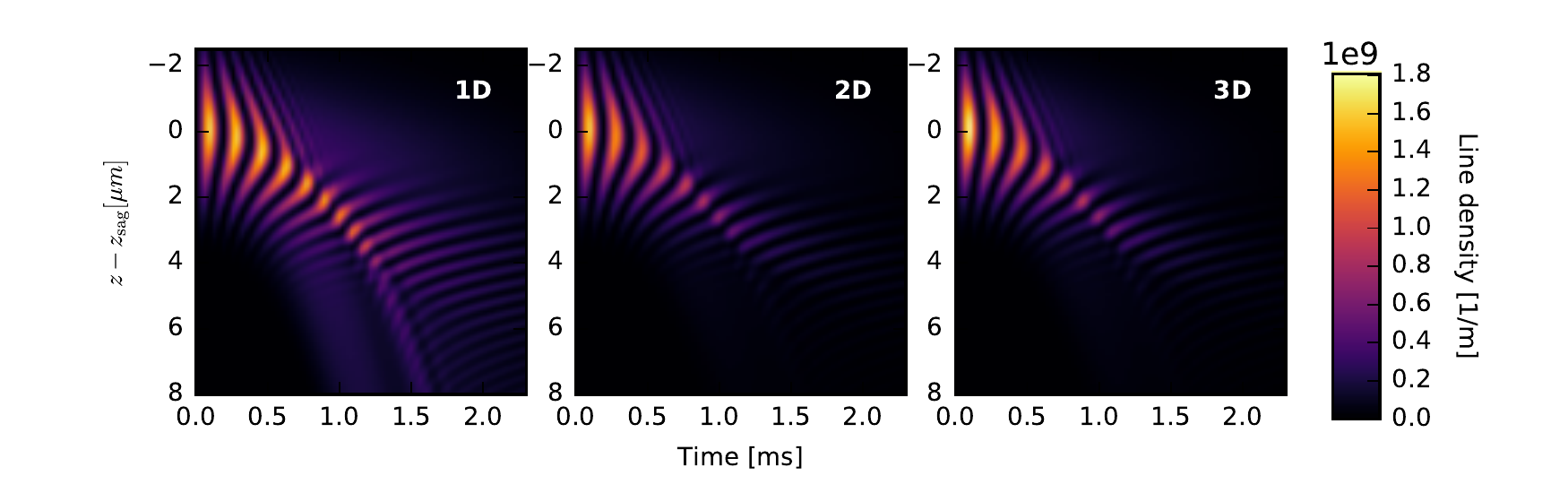}
    \caption{Comparing the integrated line density in the untrapped state along the $z$-axis (direction of gravity) obtained from 1D, 2D, and 3D simulations in the strong-coupling limit $(\Omega = \unit[2\pi\times 5.3]{kHz})$. Initially, the condensate is trapped. Outcoupling leads to the transfer of density to the unbound state that can be observed as line density below the trap. In the 2D and 3D simulations, outcoupling decreases significantly.}
    \label{fig:strong-coupling-limit-density}
\end{figure*}

The results from the numerical solution of the CGPE are compared to experimental data from a cold atom chip experiment \cite{gunther05}. Here, $^{ 87}\mathrm{Rb}$ atoms in the $\ket{F=2, m_F=2}$ ground state are magnetically trapped in a harmonic potential with trapping frequencies as given in Table~\ref{tab:experiment-parameters}. Outcoupling from the trapped state to the untrapped $\ket{F=1, m_F=1}$ state is achieved by irradiating a microwave magnetic field close to $\unit[6.8]{GHz}$. The outcoupled atoms are measured state-selectively with single atom sensitivity \cite{stibor10}. To this effect, they are ionized via a three-photon ionization process with laser beams placed $\unit[450]{\mu m}$ underneath the trap position. The ions are then guided by an ion optics onto a channel electron multiplier and detected with $\unit[8]{ns}$ time resolution. To avoid saturation of the detector at high microwave intensities, the efficiency of the single atom detection scheme has been tuned down to approximately $0.7\%$ as extracted from absorption images \cite{federsel15}. The Rabi frequency was calibrated by a Landau-Zener frequency sweep observing the remaining atom number fraction \cite{federsel15, zener32}.

Figure~\ref{fig:spectral-response} shows the spectral response. Due to the resonance conditions introduced in Sect.~\ref{sect:model:outcoupling}, the ICR varies with the detuning frequency $\Delta$. The Rabi frequency was fixed at $\Omega=\unit[2\pi\times 83]{s^{-1}}$. For the simulation, the CGPE were solved in up to 3D at pointwise fixed detuning frequencies. Quantitative agreement is reached with the 3D simulation around the point of maximum outcoupling. While the width of the rate model and the simulated data matches, the experimental data extends over a wider range. This is attributed to finite temperature, where the condensate is surrounded by a thermal cloud. Experimentally, the detuning frequency was swept through the resonant range with a rate of $\unit[1.3]{MHz/s}$. For reasons of computational efficiency, this was studied numerically only in 1D and 2D. As shown in Fig.~\ref{fig:spectral-response-sweep}, sweeping the detuning frequency yields better agreement, regarding the shape at detuning frequencies below the point of maximum outcoupling, with the output-coupling rate than the simulation with fixed detuning frequencies. 
Generally, only 3D simulations are expected to yield quantitative agreement with experimental data, as the exact shape of the ground state cannot be obtained in a lower-dimensional simulation. 

Figure~\ref{fig:sensitivity} shows the sensitivity, i.~e.~the response of the ion count rate on the Rabi frequency driving the outcoupling. For measurements in the high-intensity regime, the detection efficiency was tuned further down to $\eta = 0.007$. As expected from the rate model given in \cite{federsel15}, the ICR is proportional to $\Omega^2$ in the low-intensity regime. This is observed in both the 1D, 2D, and 3D simulations. When Rabi oscillations set in, the system enters a bound state and the atom flux decreases. The rate model is no longer applicable. For high Rabi frequencies, the 1D simulation diverges from the 2D and 3D simulations, see Fig.~\ref{fig:strong-coupling-limit-density}. This is attributed to the Rabi oscillation causing dynamics in the radial direction that was omitted in the 1D simulation. Nevertheless, the numerical simulations in all dimensions are in qualitative agreement with the experimentally observed data up to $\Omega = \unit[5\times 10^{3}]{1/s}$, whereas the semiclassical rate model is no longer applicable in the strong-coupling regime. Beyond this rate the numerical simulations deviate from the experimental data. Qualitatively, the 1D simulation yields here the best description. Note that in this regime, determining an initial count rate becomes challenging both numerically and experimentally, as the BEC is fully depleted within few ms. Thus, the out-coupling rate is not constant throughout the simulation and measurement time, respectively. From a theoretical perspective, as the BEC is depleted, the mean-field description might no longer be valid. Descriptions of ultracold, condensed gases beyond the mean-field level have been implemented \cite{schurer15a, bolsinger17}, but the combination of these approaches with the out-coupling mechanisms is left for future work. 

\section{Conclusion and Outlook}
\label{sect:conclusion}

In this paper, we have calculated out-coupling rates of an atom laser numerically by solving a set of coupled Gross-Pitaevskii equations in up to three dimensions. The rates were compared to experimental data and a rate model, and quantitative agreement was reached in the weak-coupling limit within the full three-dimensional simulation. While one-dimensional simulations provide a convenient tool to study the atom laser in a qualitative way, quantitative agreement cannot be reached, as the exact shape of the trapped ground state cannot be reproduced and radial dynamics within the condensate are not described. In the strong-coupling limit, a bound state of the atom laser is formed, where atoms are reabsorbed to the trapped state before they can leave the trap potential. This bound state is described by the 1D, 2D, and 3D simulations, and experimental data is matched qualitatively. The limitations of the mean-field description for a BEC driven by a strong out-coupling field are discussed.

\begin{acknowledgments}
  We gratefully thank the bwGRiD project for the computational resources. We gratefully acknowledge support by the Deutsche Forschungsgemeinschaft through SPP 1929 (GiRyd).
\end{acknowledgments}

\clearpage
%

\end{document}